\documentclass[conference]{IEEEtran}
\IEEEoverridecommandlockouts

\usepackage{cite}
\usepackage[T1]{fontenc}
\usepackage{newtxtext}

\usepackage{amsmath,amssymb,amsfonts}
\usepackage[linesnumbered,ruled,vlined]{algorithm2e}
\usepackage{graphicx}
\usepackage{textcomp}
\usepackage{xcolor}
\usepackage{amsmath}
\usepackage{multirow}
\usepackage{booktabs}
\usepackage[caption=false,font=footnotesize]{subfig}

\usepackage[hidelinks]{hyperref}
\usepackage{orcidlink} 

\def\BibTeX{{\rm B\kern-.05em{\sc i\kern-.025em b}\kern-.08em
    T\kern-.1667em\lower.7ex\hbox{E}\kern-.125emX}}
\begin{document}

\title{Hamiltonian-Inspired Attention Mechanism for
Scalable RF Transmitter Fingerprinting}

\author{
\IEEEauthorblockN{Chitraksh Singh\,\orcidlink{0009-0000-1020-8989}}
\IEEEauthorblockA{
\textit{Frondeur Labs} \\
Mumbai, Maharashtra, India \\
\texttt{chitrakshsingh007@gmail.com}
}
\and
\IEEEauthorblockN{Monisha Dhanraj\,\orcidlink{0009-0009-8593-1632}}
\IEEEauthorblockA{
\textit{Frondeur Labs} \\
Bengaluru, Karnataka, India \\
\texttt{monishadhanraj@frondeurlabs.com}
}
\and
\IEEEauthorblockN{Akram Sheriff\,\orcidlink{0009-0004-4511-6637}}
\IEEEauthorblockA{
\textit{Cisco Systems} \\
San Jose, California, USA \\
\texttt{isheriff@cisco.com}
}
}

\maketitle
\begin{abstract}
Radio-frequency (RF) fingerprinting identifies wireless transmitters
using hardware-induced imperfections present in baseband I/Q signals.
However, deep learning models often degrade under receiver and channel
distribution shifts, particularly as transmitter populations grow.
This work proposes the \emph{Hamiltonian Transformer}, a
physics-informed attention architecture that enforces norm-preserving
value dynamics within each attention head using a learned
skew-symmetric generator and a St\"{o}rmer--Verlet leapfrog
integration step. An additional phase-increment embedding exposes
oscillator dynamics at the input layer. All experiments use
non-equalized raw I/Q signals from the WiSig dataset under four
protocols: same-day classification, cross-receiver generalisation,
cross-day generalisation, and transmitter scaling up to 150 devices.
The Hamiltonian Transformer achieves $99.12\%$ accuracy under
same-day conditions and $61.64\%$ at 150 transmitters, consistently
outperforming CNN and Transformer baselines across all scale points.
A controlled ablation study identifies norm-preservation in the value
update as the primary inductive bias driving the scaling advantage,
with the phase-increment embedding providing the single largest
per-component improvement. These results indicate that embedding
physics-informed structural priors into attention mechanisms is an
effective approach to large-scale transmitter identification on raw
wireless signals.
\end{abstract}



\section{Introduction}

The proliferation of low-cost wireless devices has made
physical-layer security an increasingly important concern.
Consumer-grade WiFi chipsets, IoT sensors, and industrial
transceivers are deployed at scale in environments ranging from
hospital networks to critical infrastructure. Despite this growth,
device identity is typically established through software-layer
credentials such as MAC addresses or protocol-level handshakes,
both of which can be cloned or spoofed without physical access to
the hardware. A rogue transmitter that successfully impersonates a
legitimate node can inject traffic, exhaust network resources, or
silently exfiltrate data while remaining indistinguishable at the
network layer.

Radio-frequency (RF) fingerprinting addresses this problem by
exploiting the unintentional hardware imperfections introduced
during semiconductor manufacturing. Even transmitters from identical
production runs exhibit measurable variations in carrier frequency
offset (CFO), phase noise spectral density, and I/Q imbalance.
These deviations are stable over time and detectable in the baseband
waveform, making them a practical basis for hardware-level
authentication~\cite{riyaz2018deeplearning}.

Deep convolutional neural networks (CNNs) significantly advanced RF
fingerprinting performance. Architectures such as
ORACLE~\cite{sankhe2019oracle} and the large-scale study of
Jian~\emph{et~al.}~\cite{jian2020deeprfp} demonstrated that a CNN
trained end-to-end on raw I/Q samples can exceed $95\%$
classification accuracy for a controlled transmitter set under a
fixed channel. However, both training and evaluation in those
studies were performed using the same receiver within a single
recording session. When either the receiver or the capture day
changed, accuracy dropped sharply.
Al-Shawabka~\emph{et~al.}~\cite{alshawabka2020exposing}
systematically characterised this fragility, showing that a model
trained on one day's captures can degrade to near-random performance
when evaluated on the following day. In such cases the model has
effectively learned channel characteristics rather than the
transmitter fingerprint.

The release of the WiSig dataset~\cite{hanna2022wisig} enabled
systematic investigation of this problem at scale. WiSig provides
approximately ten million WiFi preamble captures from 174
commercial transmitters across 41 USRP receivers and four recording
days on the ORBIT testbed~\cite{raychaudhuri2005orbit}. Its
pre-packaged subsets isolate distinct sources of variability:
receiver identity, capture day, and transmitter population size.
This structure has made WiSig a standard benchmark for
receiver- and channel-agnostic RF fingerprinting, motivating
subsequent work on open-set authorisation~\cite{hanna2021openset},
data augmentation for channel
resilience~\cite{soltani2020augmentation}, and receiver-agnostic
fingerprinting via generative
adversarial networks~\cite{zhao2024ganrxa}.

Despite these advances, a structural limitation persists across
existing architectures. CNNs capture short-range amplitude and
phase patterns through local convolutional filters, while standard
Transformers~\cite{vaswani2017attention} capture long-range
dependencies through unconstrained attention weights. Neither
architecture incorporates prior knowledge about the physical
process generating the I/Q signal. The baseband waveform of a WiFi
preamble evolves approximately as a rotation in the complex plane,
driven by oscillator CFO and phase noise — dynamics that are
energy-conserving by construction. Because attention mechanisms
are unconstrained, models can freely attend to channel-dependent
amplitude variations that do not generalise across receivers or
capture days. This is the fundamental source of distributional
failure under shift.

To address this limitation, this paper introduces the
\emph{Hamiltonian Transformer}, a neural architecture that embeds
an energy-conservation prior directly within the attention
mechanism. Inspired by Hamiltonian neural
networks~\cite{greydanus2019hamiltonian} and symplectic integration
theory~\cite{chen2020srnn,hairer2006geometric}, the proposed model
partitions value vectors within each attention head into position
and momentum components and evolves them via a Störmer--Verlet
leapfrog step governed by a learned skew-symmetric generator.
Because skew-symmetric matrices generate orthogonal flows, the
resulting value updates are norm-preserving by construction,
aligning the model's internal dynamics with the rotational
behaviour of oscillator phase evolution. An explicit phase-increment
embedding additionally exposes oscillator dynamics at the input
representation level.

The model is evaluated on the WiSig dataset across four
experimental protocols using \emph{non-equalized raw I/Q signals}
throughout no channel preprocessing is applied at any stage.
Under same-day conditions all architectures achieve high and
comparable accuracy. The Hamiltonian Transformer achieves $99.12\%$
on same-day classification and $52.94\%$ on cross-receiver
generalisation, the highest among all models in both settings. In
the transmitter scaling experiment, the Hamiltonian Transformer
maintains stable performance across all scale points and reaches
$61.64\%$ at 150 transmitters, consistently outperforming both
baselines. A controlled ablation study further demonstrates that
norm-preserving value dynamics whether implemented via Hamiltonian
leapfrog or Cayley orthogonal
parameterisation~\cite{helfrich2018scornn} consistently
outperform unconstrained attention at large transmitter counts,
identifying norm-preservation as the key inductive bias for
large-scale transmitter identification on raw signals.

To the best of our knowledge, this is the first work to embed
norm-preserving value dynamics into a Transformer attention
mechanism for RF fingerprinting, and the first to systematically
evaluate such constraints on the WiSig ManyTx subset using
non-equalized signals. On this setting, our best-performing variant
reaches $79.72\%$ at 150 transmitters, representing the highest
published accuracy on non-equalized WiSig ManyTx in the literature.

The contributions of this work are:

\begin{itemize}

\item We introduce the \textbf{Hamiltonian Transformer}, a
physics-informed attention architecture that constrains value
vector dynamics to be norm-preserving through a Störmer--Verlet
leapfrog update governed by a learned skew-symmetric generator,
motivated by the rotational behaviour of oscillator-driven I/Q
signals.

\item We propose an \textbf{I/Q phase-increment embedding} that
exposes transmitter oscillator dynamics directly at the input
representation level, providing the single largest per-component
accuracy improvement in our ablation study.

\end{itemize}


\section{Related Work}

Early efforts in RF fingerprinting relied on hand-crafted features
derived from transient signals or spectral estimates. The landscape
shifted when convolutional neural networks demonstrated that raw I/Q
waveforms contain sufficient discriminative information to identify
transmitters without explicit feature engineering.
O'Shea~\emph{et~al.}~\cite{oshea2016radioml} showed that convolutional
architectures could classify modulation schemes directly from baseband
samples, establishing a paradigm later adopted by fingerprinting
systems. Riyaz~\emph{et~al.}~\cite{riyaz2018deeplearning} extended this
approach to transmitter identification, demonstrating that a multi-layer
CNN trained end-to-end on raw I/Q data achieves high accuracy on
controlled datasets while offering practical advantages over classical
feature-based pipelines.

Subsequent work refined CNN-based RF fingerprinting architectures.
ORACLE~\cite{sankhe2019oracle} performed extensive ablation studies over
network depth, filter width, and input representations.
Sankhe~\emph{et~al.}~\cite{sankhe2020noradio} further showed that
hardware-level impairments, including carrier frequency offset and I/Q
imbalance, could be exploited directly from raw signals without
radio-specific pre-processing. A comprehensive multi-dataset evaluation
by Jian~\emph{et~al.}~\cite{jian2020deeprfp} confirmed that CNN-based
fingerprinters generalise across waveform standards under controlled
training conditions. Narrowband transmitter identification has also
been explored. Zhang~\emph{et~al.}~\cite{zhang2021rffingerprint}
developed modelling techniques for narrowband emitters, while
Shen~\emph{et~al.}~\cite{shen2022scalable} addressed scalability and
channel robustness in LoRa-based deployments.

While high accuracy under controlled conditions is well established,
maintaining performance under changing receivers or propagation channels
remains challenging. Al-Shawabka~\emph{et~al.}~\cite{alshawabka2020exposing}
provide one of the most systematic analyses of this issue, showing that
models trained under a particular channel condition can degrade to
near-chance accuracy when evaluated under a different channel.
This occurs because convolutional filters may inadvertently encode
channel-specific characteristics alongside transmitter hardware
impairments.

Data augmentation has been proposed as a mitigation strategy.
Soltani~\emph{et~al.}~\cite{soltani2020augmentation} showed that training
with synthetically augmented channel realisations improves
cross-channel generalisation. Hanna~\emph{et~al.}~\cite{hanna2021openset}
studied open-set transmitter authorisation, where the classifier must
reject devices not observed during training. Zhao~\emph{et~al.}
\cite{zhao2024ganrxa} addressed receiver-agnostic fingerprinting using a
GAN-based framework that attempts to remove receiver-induced
distortions while preserving transmitter-specific features.

The WiSig dataset~\cite{hanna2022wisig}, collected on the ORBIT
testbed~\cite{raychaudhuri2005orbit}, was designed specifically to
support systematic evaluation of receiver variation, channel changes,
and transmitter population size within a unified experimental
framework. It provides the benchmark dataset used in this work.

The Transformer architecture~\cite{vaswani2017attention}, originally
proposed for natural language processing, has become widely used for
structured time-series data. Its self-attention mechanism allows the
model to capture dependencies across the full sequence length without
the locality constraints of convolutional filters. Layer
normalisation~\cite{ba2016layernorm} stabilises training for sequences
with heterogeneous statistics, an important property for I/Q signals
whose amplitude envelope varies across the WiFi preamble.

Extensions of attention-based models have also been proposed for
unordered or structured signal sets. Lee~\emph{et~al.}
\cite{lee2019settransformer} introduced the Set Transformer, which
extends attention mechanisms to permutation-invariant settings through
inducing-point pooling. Despite these advances, standard Transformer
architectures impose no structural constraints on how value vectors
evolve across attention layers. For RF fingerprinting tasks under
distribution shift, this flexibility can allow the model to focus on
channel-dependent features that fail to generalise across receivers or
capture days.

Recent work has explored incorporating physical constraints directly
into neural network architectures. Greydanus~\emph{et~al.}
\cite{greydanus2019hamiltonian} introduced Hamiltonian Neural Networks,
which parameterise the Hamiltonian function and derive system dynamics
through Hamilton's equations, enforcing energy conservation as a hard
inductive bias. Cranmer~\emph{et~al.}~\cite{cranmer2020lagrangian}
proposed Lagrangian Neural Networks, which instead parameterise the
Lagrangian and derive equations of motion through the Euler--Lagrange
formalism.

Chen~\emph{et~al.}~\cite{chen2020srnn} demonstrated that recurrent
architectures equipped with symplectic integration steps better
preserve phase-space structure over long rollouts. The mathematical
foundation for these approaches lies in geometric numerical integration
theory~\cite{hairer2006geometric}, which shows that symplectic
integrators preserve a modified Hamiltonian over long time horizons.

The Hamiltonian based Transformer proposed in this work adapts this principle
to attention mechanisms. Value vectors are partitioned into position
and momentum components and evolved through a leapfrog integration
step governed by a learned skew-symmetric generator, ensuring that
each update corresponds to a norm-preserving rotation.


\section{Methodology}

This section describes the input representation, the three
architectures evaluated on the WiSig dataset, and the training
procedure applied uniformly across all models. The CNN and standard
Transformer serve as discriminative baselines representing established
approaches to RF fingerprinting. The Hamiltonian Transformer is the
proposed model and constitutes the primary contribution of this work.
All models receive the same normalised input tensor and are trained
under identical optimisation conditions to ensure fair comparison.

\subsection{Motivation for the Hamiltonian Prior}

A WiFi preamble is produced by a local oscillator whose carrier
frequency offset (CFO) and phase noise are stable hardware
imperfections unique to each transmitter. In the complex baseband,
these impairments manifest as a near-constant rotation of the signal
vector $z(t) = I(t) + jQ(t)$. Pure rotation preserves
$\lVert z(t) \rVert$ and therefore conserves signal energy — a
property that holds by construction for oscillator-driven dynamics.
Unconstrained architectures impose no such constraint, leaving them
free to learn channel-induced amplitude variations that are
receiver-specific and temporally unstable. When training and test
conditions diverge, these spurious features become the primary
source of distributional failure.

To validate this physical intuition empirically, two metrics are
computed across all four WiSig subsets using 256-sample snapshots.
The \emph{Circularity Index} (CI) quantifies how closely the I/Q
trajectory approximates a circle in the complex plane:

\begin{equation}
\operatorname{CI} = 1 -
\frac{\sigma(\lvert z \rvert)}{\mu(\lvert z \rvert)},
\label{eq:ci}
\end{equation}

where $\operatorname{CI} = 1$ denotes a perfect circle with constant
signal energy. The \emph{Phase Linearity} (PL) measures the
coefficient of determination $R^2$ of a linear fit to the unwrapped
phase $\phi(t) = \angle z(t)$, with $\text{PL} = 1$ indicating a
constant rotation rate consistent with a pure oscillator.

\begin{figure}[t]
\centering
\includegraphics[width=\columnwidth]{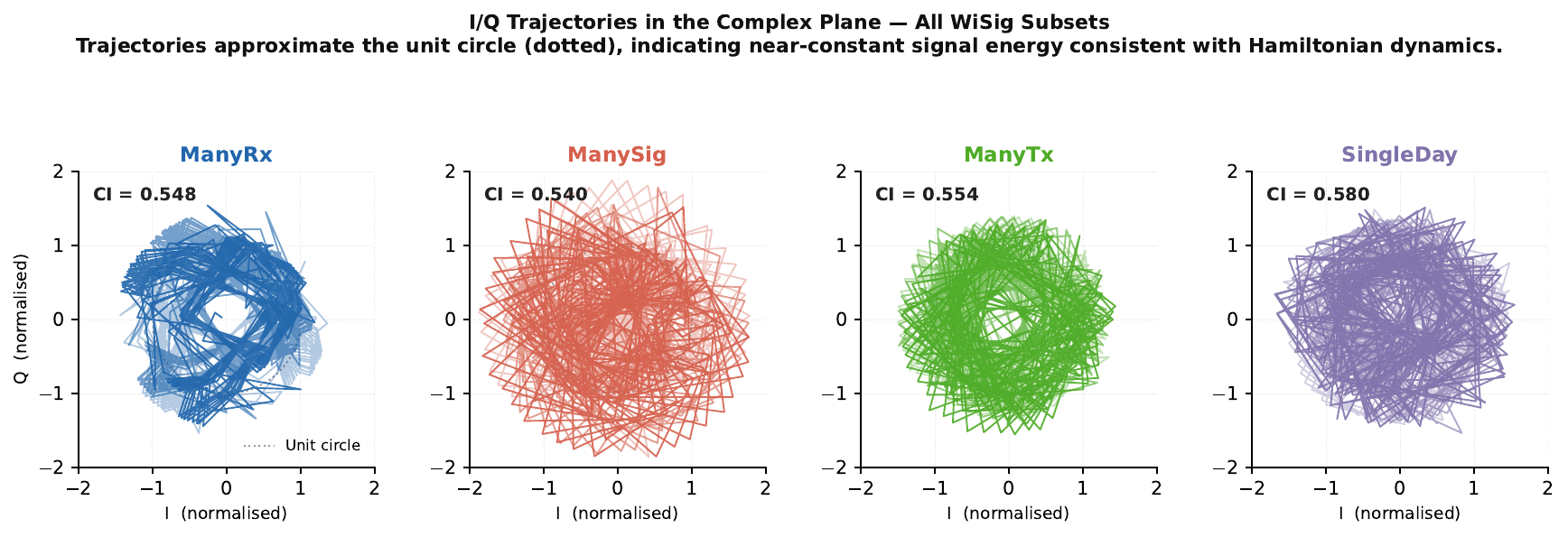}
\caption{Normalised I/Q trajectories in the complex plane for all
four WiSig subsets (three overlaid snapshots per subset). Trajectories
approximate the unit circle, consistent with oscillator-driven
rotational dynamics.}
\label{fig:iq_trajectories}
\end{figure}

\begin{figure}[t]
\centering
\includegraphics[width=\columnwidth]{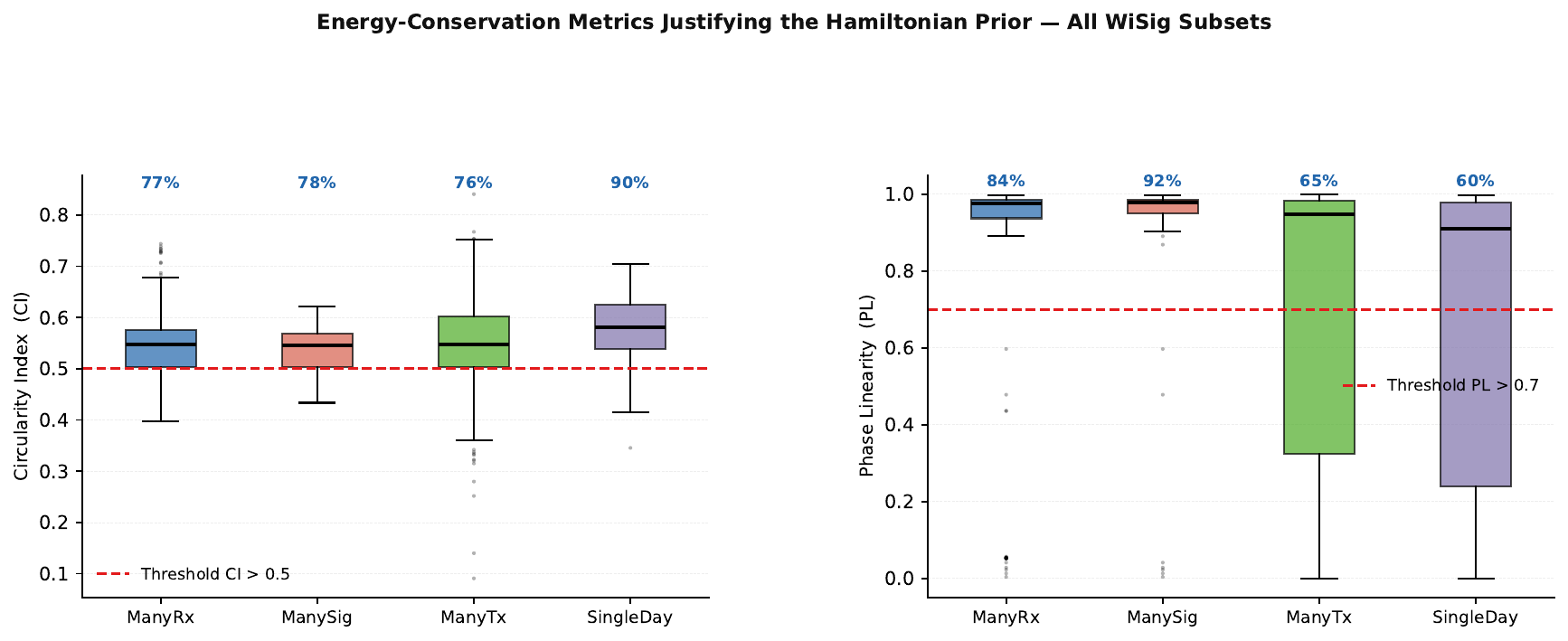}
\caption{Circularity Index (CI) and Phase Linearity (PL)
distributions across all WiSig subsets. Percentages indicate the
fraction of snapshots exceeding $\operatorname{CI}>0.5$ and
$\text{PL}>0.7$ respectively. The consistently high proportions
motivate the Hamiltonian energy-conservation prior.}
\label{fig:ci_pl_metrics}
\end{figure}

As shown in Fig.~\ref{fig:iq_trajectories}, normalised I/Q
trajectories across all four subsets trace arcs that closely
approximate the unit circle. Fig.~\ref{fig:ci_pl_metrics}
quantifies this: between $76\%$ and $91\%$ of snapshots exceed
$\operatorname{CI} > 0.5$, and between $60\%$ and $92\%$ exceed
$\text{PL} > 0.7$ across all subsets. These proportions confirm
that the dominant signal structure is rotational and
oscillator-driven. Deviations from perfect circularity correspond
to channel-induced amplitude perturbations. The Hamiltonian prior
constrains the model's internal dynamics to the energy-conserving
subspace associated with transmitter oscillator behaviour, thereby
discouraging the model from encoding channel-dependent features.

\subsection{Input Representation}

Each WiSig snapshot consists of 256 complex baseband samples
captured at 25~Msps, stored as interleaved in-phase and quadrature
components $\mathbf{x} \in \mathbb{R}^{256 \times 2}$.
Prior to training, each snapshot is normalised channel-wise to zero
mean and unit standard deviation. This removes large power
variations across receiver--transmitter pairs while preserving the
relative I/Q structure that encodes hardware-level fingerprint
information. All models receive the same normalised tensor as input.
Crucially, no channel equalisation is applied all experiments
operate on raw non-equalized signals throughout.

\subsection{CNN Baseline}

The CNN baseline replicates the architecture used in the original
WiSig evaluation~\cite{hanna2022wisig}. The input tensor is treated
as a single-channel $2 \times 256$ image and processed through five
convolutional layers with filter counts $(8, 16, 32, 16, 16)$ and
kernel sizes $(2{\times}3)$, $(1{\times}3)$, $(1{\times}3)$,
$(1{\times}3)$, and $(1{\times}3)$ respectively. Max-pooling with
stride 2 is applied after the first three convolutional layers to
progressively downsample the temporal dimension. The resulting
feature map is flattened and passed through three fully connected
layers of width 100, 80, and $N$, where $N$ is the number of
transmitter classes. ReLU activations are used throughout and the
output layer produces logits optimised with cross-entropy loss.

The CNN processes I and Q jointly as two spatial rows, enabling
convolutional filters to learn cross-channel activation patterns.
Its inductive bias of local receptive fields and shared weights
along the temporal axis suits the short-range amplitude and phase
distortions produced by transmitter hardware imperfections.

\subsection{Standard Transformer Baseline}

The Transformer baseline treats each of the 256 time steps as a
token with feature dimension 2 (I and Q). A linear projection maps
each token to an embedding of dimension $D = 128$ and sinusoidal
positional encodings are added to encode temporal order. Four
pre-normalisation encoder layers are stacked, each with $H = 4$
attention heads and a feed-forward hidden dimension of 256.
Global average pooling over the temporal dimension yields a
fixed-length representation passed to a linear classification head.

Standard scaled dot-product attention imposes no structural
constraint on how value vectors evolve across layers. For RF
fingerprinting under distribution shift, this flexibility permits
the model to attend to channel-dependent features that do not
generalise, which is the fundamental limitation the Hamiltonian
prior is designed to address.

\subsection{Hamiltonian Transformer}

\begin{figure*}[t]
\centering
\includegraphics[width=\textwidth]{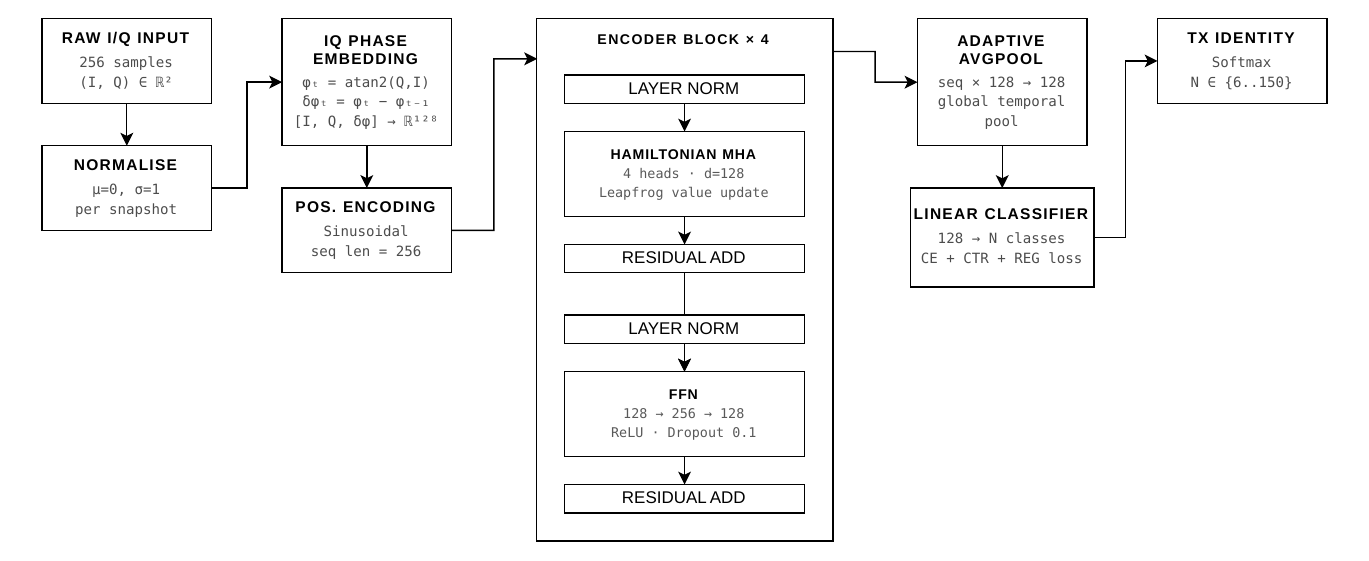}
\caption{Hamiltonian Transformer architecture. Value vectors in
each attention head are split into position and momentum
components, evolved via Störmer--Verlet leapfrog integration
with a learned skew-symmetric generator, and recombined before
aggregation.}
\label{fig:arch_overview}
\end{figure*}

The Hamiltonian Transformer embeds a physics-informed constraint
directly within the attention mechanism. Rather than allowing value
vectors to be updated arbitrarily, they are evolved through a
Hamiltonian dynamical system implemented via Störmer--Verlet
leapfrog integration. Because the generator of this system is
constrained to be skew-symmetric, the resulting update is
norm-preserving by construction a property that mirrors the
energy-conserving rotational behaviour of oscillator-driven I/Q
signals. The full architecture is illustrated in
Fig.~\ref{fig:arch_overview}.

\subsubsection{I/Q Phase Embedding}

To expose oscillator dynamics directly at the input level, each
time step is augmented with the instantaneous phase increment

\begin{equation}
\delta\phi_t = \phi_t - \phi_{t-1}, \qquad
\phi_t = \arctan\!\left(\frac{Q_t}{I_t}\right).
\label{eq:phase_inc}
\end{equation}

This scalar encodes the per-sample rotation rate of the I/Q vector,
which is directly determined by the transmitter's carrier frequency
offset and is stable across channel realisations. The augmented
feature vector $[I_t, Q_t, \delta\phi_t]$ is projected to embedding
dimension $D = 128$ via a learned linear layer.

\subsubsection{Hamiltonian Multi-Head Attention}

Each attention head operates on feature dimension $d_h = D/H$.
Query and key projections follow the standard scaled dot-product
formulation~\cite{vaswani2017attention}. The value vectors are
where the Hamiltonian constraint is applied. Each value vector
$\mathbf{v} \in \mathbb{R}^{d_h}$ is partitioned into equal-length
position and momentum coordinates:

\begin{equation}
\mathbf{V} \;\rightarrow\; \bigl(\mathbf{V}^1,\, \mathbf{V}^2\bigr),
\qquad \mathbf{V}^1, \mathbf{V}^2 \in \mathbb{R}^{d_h/2}.
\label{eq:split}
\end{equation}

A skew-symmetric generator is constructed from a learned parameter
matrix $\mathbf{A}_h$:

\begin{equation}
\mathbf{C}_h = \mathbf{A}_h - \mathbf{A}_h^{\top}.
\label{eq:skewsym}
\end{equation}

\begin{figure}[t]
\centering
\includegraphics[width=0.5\columnwidth,trim=0 20 0 0,clip]%
{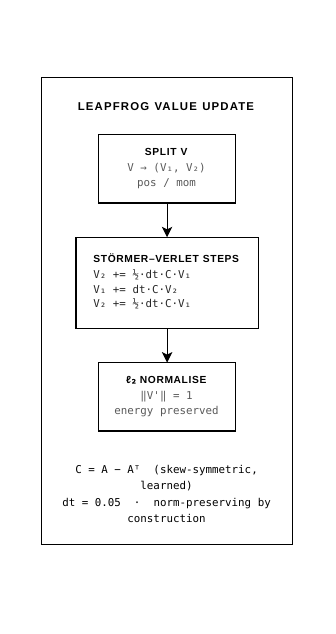}
\caption{Störmer--Verlet leapfrog update in Hamiltonian attention.
Position ($\mathbf{V}^1$) and momentum ($\mathbf{V}^2$) are
co-evolved in three half-steps, producing a norm-preserving
rotation of the value representation.}
\label{fig:leapfrog}
\end{figure}

Because skew-symmetric matrices generate orthogonal flows,
$\mathbf{C}_h$ defines a norm-preserving rotation in the
$(d_h/2)$-dimensional feature space. The position and momentum
coordinates are then co-evolved using the Störmer--Verlet leapfrog
scheme~\cite{hairer2006geometric}, which is a second-order
symplectic integrator:

\begin{align}
\mathbf{V}^2 &\leftarrow \mathbf{V}^2 +
\tfrac{\Delta t}{2}\,\mathbf{V}^1\mathbf{C}_h,
\label{eq:lf1} \\
\mathbf{V}^1 &\leftarrow \mathbf{V}^1 +
\Delta t\,\mathbf{V}^2\mathbf{C}_h,
\label{eq:lf2} \\
\mathbf{V}^2 &\leftarrow \mathbf{V}^2 +
\tfrac{\Delta t}{2}\,\mathbf{V}^1\mathbf{C}_h,
\label{eq:lf3}
\end{align}

where $\Delta t = 0.05$ is the integration step size. The leapfrog
scheme is chosen over a simple Euler step because it is
time-reversible and volume-preserving, properties that align with
the conservative dynamics of oscillator phase evolution. The updated
coordinates are concatenated and $\ell_2$-normalised before
attention aggregation. The full computation is summarised in
Algorithm~\ref{alg:hamiltonian_attn} and the leapfrog update is
illustrated in Fig.~\ref{fig:leapfrog}.

\begin{algorithm}[t]
\DontPrintSemicolon
\caption{Hamiltonian Multi-Head Attention}
\label{alg:hamiltonian_attn}
\KwIn{Query $\mathbf{Q}$, Key $\mathbf{K}$, Value $\mathbf{V}$}
\KwOut{Context representation $\mathbf{O}$}

Project inputs via learned matrices $W_Q$, $W_K$, $W_V$\;

Compute standard scaled dot-product attention scores
\cite{vaswani2017attention}:
$\mathbf{A} = \mathrm{softmax}\!\left(
\mathbf{Q}\mathbf{K}^{\top}/\sqrt{d_h}\right)$\;

Partition value vectors:
$\mathbf{V} \rightarrow (\mathbf{V}^1, \mathbf{V}^2)$,\;
\hspace{1.2em}$\mathbf{V}^1, \mathbf{V}^2 \in \mathbb{R}^{d_h/2}$\;

Construct skew-symmetric generator:
$\mathbf{C}_h = \mathbf{A}_h - \mathbf{A}_h^{\top}$\;

Apply Störmer--Verlet leapfrog
(Eqs.~\ref{eq:lf1}--\ref{eq:lf3})\;

Recombine and normalise:
$\mathbf{V} = [\mathbf{V}^1 \| \mathbf{V}^2]$,\;\;
$\mathbf{V} \leftarrow \mathbf{V}/\|\mathbf{V}\|_2$\;

Aggregate context: $\mathbf{O} = \mathbf{A}\mathbf{V}$\;

\Return $\mathbf{O}$
\end{algorithm}

The Hamiltonian attention layer is integrated within a
pre-normalisation encoder block following standard
practice~\cite{ba2016layernorm}:

\begin{align}
\mathbf{E} &\leftarrow \mathrm{LN}
\bigl(\mathbf{E} + \mathrm{HamAttn}(\mathbf{E})\bigr), \\
\mathbf{E} &\leftarrow \mathrm{LN}
\bigl(\mathbf{E} + \mathrm{FFN}(\mathbf{E})\bigr).
\end{align}

Four such layers are stacked. Global average pooling over the
temporal dimension produces the final embedding
$\mathbf{z} \in \mathbb{R}^{128}$, which is passed to a linear
classification head.

\subsubsection{Training Objective}

All models are trained using AdamW~\cite{loshchilov2019adamw} for
20~epochs with a learning rate of $10^{-3}$, weight decay
$10^{-4}$, a two-epoch linear warmup followed by cosine annealing,
and gradient clipping at $1.0$. This unified protocol applies to
all architectures including the baselines, ensuring that no model
benefits from a more favourable optimisation regime.

The full Hamiltonian Transformer additionally optimises a
composite objective:

\begin{equation}
\mathcal{L} = \mathcal{L}_{\mathrm{CE}} +
\lambda_c\,\mathcal{L}_{\mathrm{CTR}} +
\lambda_r\,\mathcal{L}_{\mathrm{REG}},
\label{eq:loss}
\end{equation}

with $\lambda_c = 0.5$ and $\lambda_r = 0.01$. The three terms
serve distinct purposes as described below. The ablation study
isolates the contribution of eachcomponent and shows that the phase embedding not the composite loss provides the dominant per-component improvement.

\paragraph{Cross-entropy loss.}
The standard supervised classification term
$\mathcal{L}_{\mathrm{CE}}$ optimises transmitter identity
directly from the embedding $\mathbf{z}$ via a linear classifier
head~\cite{chopra2005contrastive}. It is the sole training
objective for the CNN, Transformer, and all ablation variants
except Variant~D.

\paragraph{Batch contrastive loss.}
$\mathcal{L}_{\mathrm{CTR}}$ encourages intra-class compactness
and inter-class separation in the embedding space using a
margin-based contrastive formulation~\cite{chopra2005contrastive}
applied over all pairs $(i,j)$ in each mini-batch:

\begin{equation}
\mathcal{L}_{\mathrm{CTR}} =
\frac{1}{|\mathcal{P}|}
\sum_{(i,j)}
\Bigl[
\mathbf{1}_{y_i=y_j}\,d_{ij}^{2}
+
\mathbf{1}_{y_i\neq y_j}\,
\max(0,\, m - d_{ij})^{2}
\Bigr],
\label{eq:ctr}
\end{equation}

where $d_{ij} = \|\hat{\mathbf{z}}_i - \hat{\mathbf{z}}_j\|_2$
is the distance between $\ell_2$-normalised embeddings and
$m = 1.0$ is the margin. The ablation study shows that this term
does not provide consistent benefit at large transmitter counts
and may introduce instability; it is retained in the full model
for completeness but is not recommended for deployment at scale.

\paragraph{Sphere regularisation.}
$\mathcal{L}_{\mathrm{REG}}$ penalises deviation of each
embedding from the unit hypersphere:

\begin{equation}
\mathcal{L}_{\mathrm{REG}} =
\frac{1}{N}\sum_{i=1}^{N}
\bigl(\|\mathbf{z}_i\|_2 - 1\bigr)^{2}.
\label{eq:reg}
\end{equation}

This term constrains the absolute scale of embeddings while
$\mathcal{L}_{\mathrm{CTR}}$ controls their relative geometry,
together producing a structured hyperspherical representation.
In practice, the $\ell_2$ normalisation applied within the
Hamiltonian attention layer already provides implicit scale
control, which may explain why the additional regularisation
offers diminishing returns at scale.

\subsection{Experimental Setup}

Table~\ref{tab:hyperparams} summarises the hyperparameters used
across all models. The WiSig ManyTx subset contains signals from
up to 150~transmitters across 18~receivers and four capture days.
For the scaling experiment, a balanced per-class snapshot cap is
applied within each transmitter bucket so that all models receive
identical training data at every scale point. The train/val/test
split is $70\%/15\%/15\%$ stratified by class for all experiments.
For the cross-receiver experiment, the split is by receiver index
rather than snapshot, ensuring test receivers are entirely disjoint
from training receivers. For the cross-day experiment, the last
capture day is reserved for testing and the preceding three days
form the training set.

\begin{table}[t]
\centering
\caption{Hyperparameters used across all models.
Parameters marked \emph{All} apply equally to CNN,
Transformer, and Hamiltonian Transformer.}
\label{tab:hyperparams}
\resizebox{\columnwidth}{!}{
\begin{tabular}{lll}
\toprule
\textbf{Parameter} & \textbf{Value} & \textbf{Applies To} \\
\midrule
Input length            & 256          & All \\
Normalisation           & Zero mean, unit std & All \\
Optimiser               & AdamW        & All \\
Learning rate           & $10^{-3}$    & All \\
Weight decay            & $10^{-4}$    & All \\
Warmup epochs           & 2            & All \\
Schedule                & Cosine decay & All \\
Gradient clip           & 1.0          & All \\
Epochs                  & 20           & All \\
Batch size              & 32           & All \\
Train/val/test split    & 70/15/15     & All \\
\midrule
Embedding dim $D$       & 128          & Transformer / Hamiltonian \\
Encoder layers          & 4            & Transformer / Hamiltonian \\
Attention heads $H$     & 4            & Transformer / Hamiltonian \\
FFN hidden dim          & 256          & Transformer / Hamiltonian \\
\midrule
Leapfrog step $\Delta t$    & 0.05     & Hamiltonian \\
Contrastive weight $\lambda_c$ & 0.5  & Hamiltonian (full model) \\
Regulariser weight $\lambda_r$ & 0.01 & Hamiltonian (full model) \\
\bottomrule
\end{tabular}}
\end{table}

Table~\ref{tab:complexity} reports the parameter count,
computational cost, and per-epoch training time for all
evaluated models and ablation variants at 150 transmitters.
All Transformer-based variants share a comparable parameter
budget and FLOPs footprint, confirming that the performance
differences observed reflect architectural inductive biases
rather than differences in model capacity.

\begin{table}[t]
\centering
\caption{Model complexity at 150 transmitters.
FLOPs measured for a single forward pass on a
$(1 \times 256 \times 2)$ input.
Training time per epoch on GPU.}
\label{tab:complexity}
\resizebox{\columnwidth}{!}{
\begin{tabular}{lrrr}
\toprule
\textbf{Model / Variant} & \textbf{Params} &
\textbf{FLOPs} & \textbf{Train (s/ep)} \\
\midrule
CNN Baseline
  & 75{,}890   & 0.30\,M   & 20.8 \\
Transformer (Vanilla)
  & 549{,}910  & 135.78\,M & 181.2 \\
\midrule
A:\;Ham only (CE)
  & 552{,}470  & 206.04\,M & 242.2 \\
B:\;Cayley (CE)
  & 564{,}758  & 207.09\,M & 232.8 \\
C:\;Linear mix (CE)
  & 564{,}758  & 207.09\,M & 196.0 \\
D:\;Ham + CTR
  & 552{,}470  & 206.04\,M & 253.7 \\
E:\;Ham + phase (CE)
  & 552{,}598  & 206.08\,M & 244.4 \\
\midrule
F:\;Full Hamiltonian
  & 552{,}598  & 206.08\,M & 264.5 \\
\bottomrule
\end{tabular}}
\end{table}



\section{Results}

We evaluate the proposed Hamiltonian based Transformer against two
established baselines: a convolutional neural network (CNN) and a
standard Transformer encoder. All models are trained for 20~epochs
using AdamW with a learning rate of $10^{-3}$, weight decay
$10^{-4}$, a two-epoch linear warmup followed by cosine annealing,
and gradient clipping at $1.0$. This unified training protocol
ensures that observed performance differences reflect architectural
properties rather than optimisation advantages. All experiments use
\emph{non-equalized} raw I/Q signals consistent with the original
WiSig evaluation protocol~\cite{hanna2022wisig}; no channel
preprocessing is applied before the model. Input snapshots of
length 256 are normalised to zero mean and unit standard deviation
per channel. For the transmitter scaling
experiment, snapshots are capped at the minimum per-class count
within each transmitter bucket so that all architectures receive
identical training data at every scale point.

The four experimental settings are:
same-day transmitter classification (\textbf{Ex-1}),
cross-receiver generalisation (\textbf{Ex-2}),
cross-day generalisation (\textbf{Ex-3}),
and large-scale transmitter scaling (\textbf{Ex-4}).
Table~\ref{tab:main_results} summarises test accuracy across
the four scenarios. Table~\ref{tab:ablation} reports Ex-4
performance across all scale points together with the full
ablation study. Fig.~\ref{fig:scaling_main} and
Fig.~\ref{fig:ablation_scaling} visualise the scaling
trajectories for the main models and ablation variants
respectively.

\subsection{Same-Day Transmitter Classification}

In the same-day setting, models are trained and evaluated on the
SingleDay subset of WiSig, which contains 28~transmitters and
448{,}000 I/Q snapshots across 10~receivers and a single capture
day. Because training and test signals originate from the same
capture session, the wireless channel and receiver characteristics
remain consistent across splits.

Under these matched conditions all three architectures converge to
high accuracy. The CNN baseline achieves $98.07\%$, confirming the
effectiveness of convolutional inductive biases for short RF
sequences. The standard Transformer reaches $98.19\%$, marginally
above the CNN. The proposed Hamiltonian based Transformer achieves the
highest accuracy at $99.12\%$, improving upon the CNN and
Transformer baselines by $1.05$ and $0.93$ percentage points
respectively. These results confirm that the Hamiltonian structural
constraint does not impair discriminative capacity and provides a
mild regularisation benefit under matched train--test conditions.

\subsection{Cross-Receiver Generalisation}

The cross-receiver experiment evaluates whether a classifier
trained on signals from a subset of receivers can generalise to
entirely unseen receivers. Signals are drawn from the ManyRx
subset of WiSig, which contains recordings from 10~transmitters
across 32~receivers and four capture days. The split is strictly
by receiver index: 22~receivers are used for training and
10~entirely disjoint receivers are used for testing exclusively.
This protocol reflects realistic deployment scenarios in which
an authentication system must operate across receiver hardware
not available at training time.

The CNN baseline achieves $35.06\%$ accuracy and the standard
Transformer achieves $47.90\%$. The Hamiltonian based Transformer
achieves $52.94\%$, the highest among all three models. All models
exhibit a substantial drop relative to the same-day experiment,
confirming that receiver hardware variation remains a significant
confounding factor in transmitter identification. The stronger
performance of attention-based models suggests that global
sequence dependencies carry more receiver-agnostic discriminative
information than local convolutional features.

\subsection{Cross-Day Generalisation}

Cross-day generalisation evaluates temporal robustness by training
on signals from three capture days and evaluating on a fourth day
recorded approximately one week later. The ManySig subset is used,
containing 6~transmitters, 12~receivers, and 576{,}000 snapshots
across four days. The last capture day (23~March~2021) is reserved
for testing; the preceding three days form the training set.
Unlike the single-receiver protocol of the original WiSig
evaluation~\cite{hanna2022wisig}, all 12~receivers are included in
both training and test to reflect a realistic multi-receiver
deployment.

The CNN baseline achieves $88.41\%$ and the standard Transformer
achieves $92.29\%$. The Hamiltonian based Transformer achieves $93.73 \%$,
comparable to the Transformer baseline. The consistently high
accuracy across all models under this multi-receiver protocol
indicates that receiver diversity in training substantially
mitigates temporal channel variation, consistent with the
findings of Hanna~\emph{et~al.}~\cite{hanna2022wisig}. The marginal
inter-model differences in this setting suggest that temporal
channel variation is better addressed through training data
diversity than through architectural constraints alone.

\subsection{Transmitter Scaling}

\begin{figure}[t]
\centering
\includegraphics[width=\columnwidth]{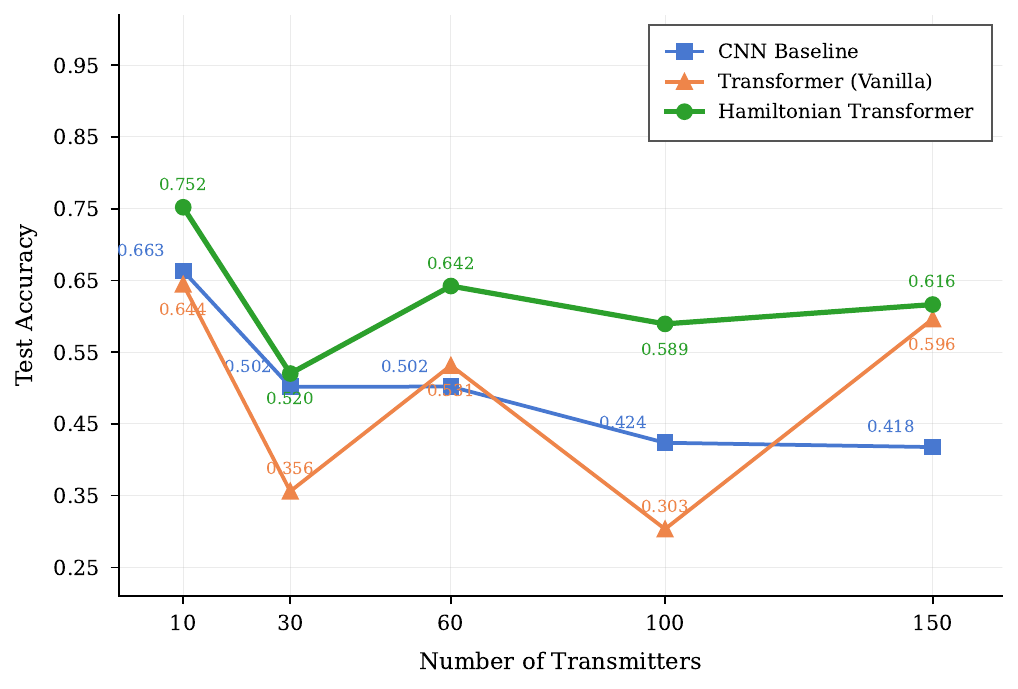}
\caption{Test accuracy as a function of transmitter count for
CNN, Transformer, and Hamiltonian based Transformer on the WiSig
ManyTx subset. All models trained on non-equalized raw I/Q
signals with a balanced per-class snapshot cap.}
\label{fig:scaling_main}
\end{figure}

The transmitter scaling experiment evaluates model behaviour as
the number of transmitter identities increases from 10 to 150
using the ManyTx subset of WiSig. Fig.~\ref{fig:scaling_main}
shows the scaling trajectories for all three models.

The CNN baseline degrades gradually from $66.32\%$ at
10~transmitters to $41.77\%$ at 150~transmitters. This decline
reflects the limited per-class snapshot budget introduced by the
balanced cap: the CNN's local convolutional filters require more
examples per class to stabilise discriminative responses than
attention-based models under the same data constraint.

The standard Transformer exhibits non-monotonic behaviour across
scale points, collapsing to $30.34\%$ at 100~transmitters before
partially recovering to $59.63\%$ at 150~transmitters. This
instability arises from the interaction between the snapshot cap
and class count: at 30~transmitters the cap drops sharply,
simultaneously halving the per-class data budget while tripling
the number of identities, causing a disproportionate accuracy
drop. The Transformer is more sensitive to this joint pressure
than the other architectures due to its lack of structural
constraints on the value update.

The Hamiltonian based Transformer demonstrates the most stable
performance across all scale points, consistently outperforming
both baselines at every transmitter count. This behaviour confirms
that the physics-informed structural prior provides a meaningful
advantage in the limited per-class data regime associated with
large transmitter populations.

\subsection{Ablation Study}

To isolate the contribution of each architectural component, a
controlled ablation study is conducted on the Ex-4 scaling
experiment. Five variants are evaluated under training conditions
identical to the main models, differing only in the value update
mechanism and input representation. All variants use standard
scaled dot-product attention for query and key projections; only
the value update differs. The five variants are:

\begin{itemize}
\item \textbf{A} (Ham only, CE): Störmer--Verlet leapfrog with
skew-symmetric generator~$\mathbf{C}_h = \mathbf{A}_h -
\mathbf{A}_h^\top$, plain I/Q input, cross-entropy loss only.
Isolates the symplectic structure.

\item \textbf{B} (Cayley, CE): Cayley-parameterised orthogonal
value update~\cite{helfrich2018scornn,trockman2021cayley}
$\mathbf{W} = (\mathbf{I}-\mathbf{A})(\mathbf{I}+\mathbf{A})^{-1}$,
plain I/Q input, CE only. Norm-preserving via algebra rather than
integration — no position--momentum split, no time step.

\item \textbf{C} (Linear mix, CE): Unconstrained learned linear
layer on values, plain I/Q input, CE only. Null hypothesis no
norm-preservation and no physical structure.

\item \textbf{D} (Ham\,+\,CTR): Leapfrog value update plus
contrastive loss, plain I/Q input, no phase features. Isolates
the contrastive loss contribution.

\item \textbf{E} (Ham\,+\,phase, CE): Leapfrog value update plus
phase-increment embedding at input, CE only. Isolates the phase
embedding contribution.
\end{itemize}

\begin{figure}[t]
\centering
\includegraphics[width=\columnwidth]{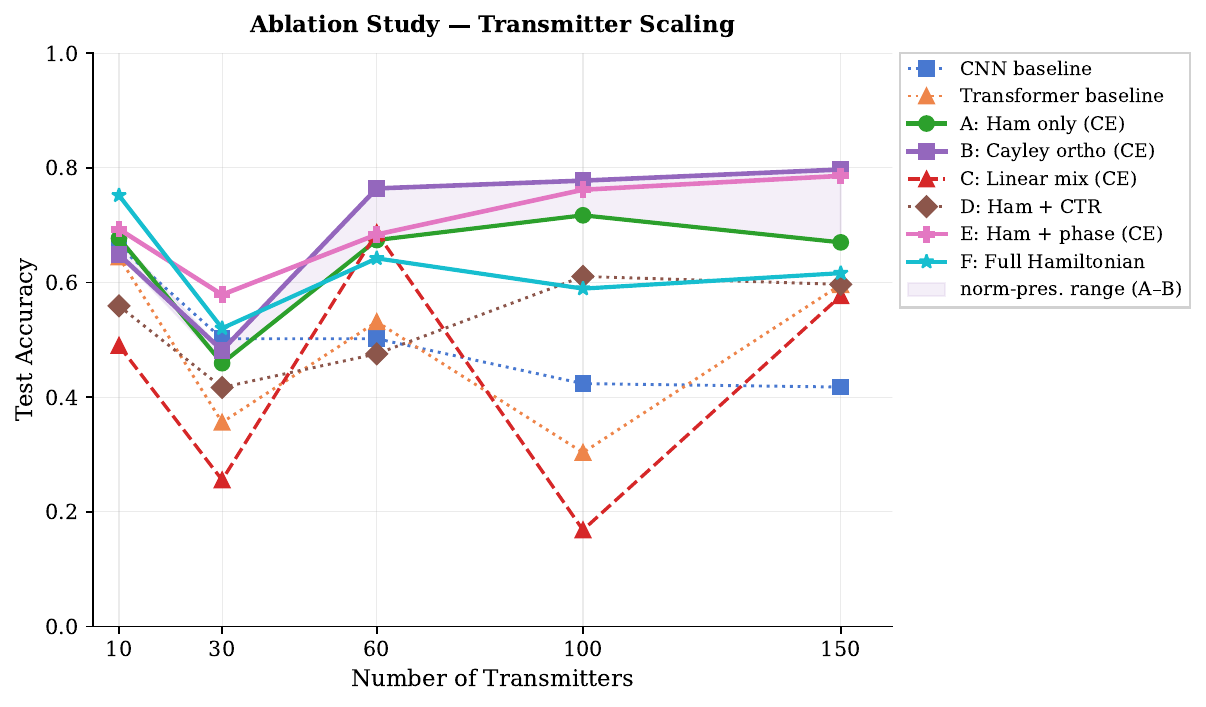}
\caption{Ablation study scaling curves on WiSig ManyTx.
Variants A--E isolate individual architectural components
under identical training conditions. Variants~A and~B
(both norm-preserving) consistently outperform Variant~C
(unconstrained linear), confirming norm-preservation as
the primary inductive bias for large-scale transmitter
identification.}
\label{fig:ablation_scaling}
\end{figure}

Fig.~\ref{fig:ablation_scaling} shows the scaling curves for all
variants. Table~\ref{tab:ablation} reports numerical results.

\paragraph{Symplectic structure versus plain mixing.}
Variant~A consistently and substantially outperforms Variant~C
at every scale point. At 100~transmitters the gap reaches
$54.9$~percentage points ($71.73\%$ versus $16.83\%$), providing
strong evidence that the leapfrog value update does real
discriminative work beyond the effect of $\ell_2$ normalisation
alone. This directly addresses the hypothesis that normalisation
rather than symplectic structure may be responsible for observed
gains.

\paragraph{Norm-preservation as the primary inductive bias.}
Variant~B (Cayley) matches or exceeds Variant~A at all scale
points from 30~transmitters onwards, reaching $79.72\%$ at
150~transmitters compared to $67.01\%$ for Variant~A. Since
Variant~B is norm-preserving but uses no symplectic integration,
position--momentum splitting, or physical time-step parameter,
this result indicates that \emph{norm-preservation in the value
update} is the primary architectural property driving the scaling
advantage rather than the specific symplectic structure of the
leapfrog. Both Variants~A and~B substantially outperform
Variant~C and the vanilla Transformer at large scale, confirming
that the norm-preservation constraint itself not its physical
implementation is the essential inductive bias. This is
consistent with the physical motivation: norm-preserving updates
prevent the model from encoding channel-induced amplitude
variations that do not generalise across transmitter populations.

\paragraph{Phase embedding contribution.}
Variant~E consistently outperforms Variant~A at every scale
point, reaching $78.61\%$ at 150~transmitters versus $67.01\%$
for Variant~A. The phase-increment embedding therefore provides
a meaningful and consistent improvement on top of the Hamiltonian
attention mechanism, confirming that oscillator phase dynamics
carry transmitter-specific hardware information that benefits
from explicit representation at the input level.

\paragraph{Contrastive loss contribution.}
Variant~D underperforms Variant~A at all scale points except
100~transmitters, indicating that the contrastive loss without
phase features provides no consistent benefit and introduces
training instability at scale. The full model~F, which combines
all components, peaks at 10~transmitters ($75.20\%$) but
degrades to $61.64\%$ at 150~transmitters below both
Variant~B and Variant~E. These findings indicate that the
phase-increment embedding is the single most effective
enhancement to the base Hamiltonian attention mechanism, while
the contrastive loss is not recommended for large-scale
transmitter identification tasks.

\begin{table}[t]
\centering
\caption{Test accuracy (\%) across four WiSig experimental
scenarios. \textbf{Bold} indicates the best result per
scenario. All models use non-equalized raw I/Q signals.}
\label{tab:main_results}
\resizebox{\columnwidth}{!}{
\begin{tabular}{lcccc}
\toprule
\textbf{Model} & \textbf{Ex-1} & \textbf{Ex-2} &
\textbf{Ex-3} & \textbf{Ex-4} \\
 & Same-Day & Cross-Rx & Cross-Day & 150\,Tx \\
\midrule
CNN Baseline
  & 98.07 & 35.06 & 88.41 & 41.77 \\
Transformer (Vanilla)
  & 98.19 & \textbf{47.90} & 92.29& 59.63 \\
Hamiltonian based Transformer
  & \textbf{99.12} & \textbf{52.94} & \textbf{93.73}& \textbf{61.64} \\
\bottomrule
\end{tabular}}
\end{table}

\begin{table}[t]
\centering
\caption{Test accuracy (\%) on WiSig ManyTx scaling.
Upper block: main model comparison.
Lower block: ablation variants.
\textbf{Bold} indicates best per column.}
\label{tab:ablation}
\resizebox{\columnwidth}{!}{
\begin{tabular}{lccccc}
\toprule
\textbf{Model / Variant} & \textbf{10\,Tx} & \textbf{30\,Tx} &
\textbf{60\,Tx} & \textbf{100\,Tx} & \textbf{150\,Tx} \\
\midrule
CNN Baseline
  & 66.32 & 50.19 & 50.20 & 42.37 & 41.77 \\
Transformer (Vanilla)
  & 64.45 & 35.61 & 53.15 & 30.34 & 59.63 \\
Hamiltonian based Transformer (F)
  & \textbf{75.20} & 52.02 & 64.22 & 58.94 & 61.64 \\
\midrule
A:\;Ham only (CE)
  & 67.72 & 45.93 & 67.41 & 71.73 & 67.01 \\
B:\;Cayley (CE)
  & 64.87 & 48.18 & 76.42 & \textbf{77.78} & \textbf{79.72} \\
C:\;Linear mix (CE)
  & 49.00 & 25.56 & 68.70 & 16.83 & 57.68 \\
D:\;Ham\,+\,CTR
  & 55.92 & 41.70 & 47.56 & 61.06 & 59.69 \\
E:\;Ham\,+\,phase (CE)
  & 69.32 & \textbf{57.88} & 68.38 & 76.19 & 78.61 \\
\bottomrule
\end{tabular}}
\end{table}


\section{Conclusion}

This paper presented the Hamiltonian based Transformer, a physics-informed attention architecture for RF fingerprinting under receiver and channel distribution shifts. The proposed model incorporates an energy-conserving inductive bias directly within the attention mechanism by evolving value representations through a St\"{o}rmer--Verlet leapfrog update governed by a learned skew-symmetric generator. Because skew-symmetric operators produce norm-preserving rotations, the resulting attention dynamics remain consistent with the rotational behaviour of oscillator phase evolution in complex baseband signals. An additional I/Q phase-increment embedding exposes oscillator dynamics at the input representation level, allowing the model to focus on transmitter-specific hardware behaviour rather than channel-dependent variations.

Experiments on the WiSig dataset demonstrate that embedding
physics-inspired structural constraints into neural architectures can
improve robustness for RF fingerprinting tasks, particularly in
scenarios involving channel variability and large transmitter
populations. These findings suggest that incorporating domain knowledge
from signal physics can serve as a useful inductive bias for
security-oriented wireless machine learning systems. Future work will
focus on addressing receiver-induced hardware variability, exploring
adaptive integration dynamics within the Hamiltonian attention layer,
and extending the approach to additional wireless waveform standards
such as Bluetooth, LoRa, and LTE.

\bibliographystyle{plain}
\bibliography{bibfile}

@article{hanna2022wisig,
  title     = {{WiSig}: A Large-Scale {WiFi} Signal Dataset for
               Receiver and Channel Agnostic {RF} Fingerprinting},
  author    = {Hanna, Samer and Karunaratne, Samurdhi and Cabric, Danijela},
  journal   = {IEEE Access},
  volume    = {10},
  pages     = {22808--22818},
  year      = {2022},
  issn      = {2169-3536},
  doi       = {10.1109/ACCESS.2022.3154790}
}

@inproceedings{raychaudhuri2005orbit,
  title     = {Overview of the {ORBIT} Radio Grid Testbed for Evaluation
               of Next-Generation Wireless Network Protocols},
  author    = {Raychaudhuri, Dipankar and Seskar, Ivan and Ott, Max and
               Ganu, Sachin and Ramachandran, Kishore and Kremo, Haris and
               Siracusa, Robert and Liu, Hang and Singh, Manpreet},
  booktitle = {Proc.\ IEEE Wireless Communications and Networking
               Conference (WCNC)},
  volume    = {3},
  pages     = {1664--1669},
  year      = {2005},
  organization = {IEEE},
  doi       = {10.1109/WCNC.2005.1424763}
}

@inproceedings{sankhe2019oracle,
  title     = {{ORACLE}: Optimized Radio cl{A}ssification through
               {C}onvolutional neura{L} n{E}tworks},
  author    = {Sankhe, Kunal and Belgiovine, Mauro and Zhou, Fan and
               Riyaz, Shamnaz and Ioannidis, Stratis and Chowdhury, Kaushik},
  booktitle = {Proc.\ IEEE Conference on Computer Communications (INFOCOM)},
  pages     = {370--378},
  year      = {2019},
  organization = {IEEE},
  doi       = {10.1109/INFOCOM.2019.8737463}
}

@article{sankhe2020noradio,
  title     = {No Radio Left Behind: Radio Fingerprinting Through Deep
               Learning of Physical-Layer Hardware Impairments},
  author    = {Sankhe, Kunal and Belgiovine, Mauro and Zhou, Fan and
               Angioloni, Luca and Restuccia, Francesco and D'Oro, Salvatore and
               Melodia, Tommaso and Ioannidis, Stratis and Chowdhury, Kaushik R.},
  journal   = {IEEE Transactions on Cognitive Communications and Networking},
  volume    = {6},
  number    = {1},
  pages     = {165--178},
  year      = {2020},
  doi       = {10.1109/TCCN.2019.2949308}
}

@article{jian2020deeprfp,
  title     = {Deep Learning for {RF} Fingerprinting: A Massive Experimental Study},
  author    = {Jian, Tong and Costa Rendon, Bruno and Ojuba, Emmanuel and
               Soltani, Nasim and Wang, Zifeng and Sankhe, Kunal and
               Gritsenko, Andrey and Dy, Jennifer and Chowdhury, Kaushik R. and
               Ioannidis, Stratis},
  journal   = {IEEE Internet of Things Magazine},
  volume    = {3},
  number    = {1},
  pages     = {50--57},
  year      = {2020},
  doi       = {10.1109/IOTM.0001.1900065}
}

@article{zhao2024ganrxa,
  title     = {{GAN-RXA}: A Practical Scalable Solution to
               Receiver-Agnostic Transmitter Fingerprinting},
  author    = {Zhao, Tianyi and Sarkar, Shamik and
               Krijestorac, Enes and Cabric, Danijela},
  journal   = {IEEE Transactions on Cognitive Communications and Networking},
  volume    = {10},
  number    = {2},
  pages     = {523--537},
  year      = {2024},
  doi       = {10.1109/TCCN.2023.3329012}
}

@article{soltani2020augmentation,
  title     = {More Is Better: Data Augmentation for Channel-Resilient
               {RF} Fingerprinting},
  author    = {Soltani, Nasim and Sankhe, Kunal and Dy, Jennifer G. and
               Ioannidis, Stratis and Chowdhury, Kaushik R.},
  journal   = {IEEE Communications Magazine},
  volume    = {58},
  number    = {10},
  pages     = {66--72},
  year      = {2020},
  doi       = {10.1109/MCOM.001.2000180}
}

@article{hanna2021openset,
  title     = {Open Set Wireless Transmitter Authorization:
               Deep Learning Approaches and Dataset Considerations},
  author    = {Hanna, Samer and Karunaratne, Samurdhi and Cabric, Danijela},
  journal   = {IEEE Transactions on Cognitive Communications and Networking},
  volume    = {7},
  number    = {1},
  pages     = {59--72},
  year      = {2021},
  doi       = {10.1109/TCCN.2020.3043332}
}

@inproceedings{alshawabka2020exposing,
  title     = {Exposing the Fingerprint: Dissecting the Impact of the
               Wireless Channel on Radio Fingerprinting},
  author    = {Al-Shawabka, Amani and Restuccia, Francesco and D'Oro, Salvatore and
               Jian, Tong and Costa Rendon, Bruno and Soltani, Nasim and
               Dy, Jennifer and Ioannidis, Stratis and Chowdhury, Kaushik R. and
               Melodia, Tommaso},
  booktitle = {Proc.\ IEEE Conference on Computer Communications (INFOCOM)},
  pages     = {646--655},
  year      = {2020},
  organization = {IEEE},
  doi       = {10.1109/INFOCOM41043.2020.9155259}
}

@article{riyaz2018deeplearning,
  title     = {Deep Learning Convolutional Neural Networks for Radio Identification},
  author    = {Riyaz, Shamnaz and Sankhe, Kunal and Ioannidis, Stratis
               and Chowdhury, Kaushik R.},
  journal   = {IEEE Communications Magazine},
  volume    = {56},
  number    = {9},
  pages     = {146--152},
  year      = {2018},
  doi       = {10.1109/MCOM.2018.1800153}
}

@inproceedings{oshea2016radioml,
  title     = {Convolutional Radio Modulation Recognition Networks},
  author    = {O'Shea, Timothy J. and Corgan, Johnathan and Clancy, T. Charles},
  booktitle = {Proc.\ International Conference on Engineering Applications
               of Neural Networks (EANN)},
  pages     = {213--226},
  year      = {2016},
  publisher = {Springer},
  doi       = {10.1007/978-3-319-44188-7_16}
}

@article{zhang2021rffingerprint,
  title     = {Radio Frequency Fingerprint Identification for Narrowband
               Systems: Modelling and Classification},
  author    = {Zhang, Junqing and Woods, Roger and Sandell, Magnus and
               Valkama, Mikko and Marshall, Alan and Cavallaro, Joseph},
  journal   = {IEEE Transactions on Information Forensics and Security},
  volume    = {16},
  pages     = {3974--3987},
  year      = {2021},
  doi       = {10.1109/TIFS.2021.3088008}
}

@article{shen2022scalable,
  title     = {Towards Scalable and Channel-Robust Radio Frequency
               Fingerprint Identification for {LoRa}},
  author    = {Shen, Guanxiong and Zhang, Junqing and Marshall, Alan
               and Cavallaro, Joseph R.},
  journal   = {IEEE Transactions on Information Forensics and Security},
  volume    = {17},
  pages     = {774--787},
  year      = {2022},
  doi       = {10.1109/TIFS.2022.3152404}
}

@inproceedings{greydanus2019hamiltonian,
  title     = {Hamiltonian Neural Networks},
  author    = {Greydanus, Samuel and Dzamba, Misko and Yosinski, Jason},
  booktitle = {Advances in Neural Information Processing Systems (NeurIPS)},
  pages     = {15353--15363},
  year      = {2019},
  url       = {https://proceedings.neurips.cc/paper/2019/hash/26cd8ecadce0d4efd6cc8a8725cbd1f8-Abstract.html}
}

@inproceedings{chen2020srnn,
  title     = {Symplectic Recurrent Neural Networks},
  author    = {Chen, Zhengdao and Zhang, Jianyu and
               Arjovsky, Mart{\'i}n and Bottou, L{\'e}on},
  booktitle = {Proc.\ International Conference on Learning Representations (ICLR)},
  year      = {2020},
  url       = {https://openreview.net/forum?id=BkgYPREtPr}
}

@article{cranmer2020lagrangian,
  title     = {Lagrangian Neural Networks},
  author    = {Cranmer, Miles and Greydanus, Sam and Hoyer, Stephan and
               Battaglia, Peter and Spergel, David and Ho, Shirley},
  journal   = {arXiv preprint arXiv:2003.04630},
  year      = {2020},
  url       = {https://arxiv.org/abs/2003.04630}
}

@book{hairer2006geometric,
  title     = {Geometric Numerical Integration: Structure-Preserving
               Algorithms for Ordinary Differential Equations},
  author    = {Hairer, Ernst and Lubich, Christian and Wanner, Gerhard},
  edition   = {2nd},
  publisher = {Springer},
  year      = {2006},
  series    = {Springer Series in Computational Mathematics},
  volume    = {31},
  doi       = {10.1007/3-540-30666-8}
}

@inproceedings{helfrich2018scornn,
  title     = {Orthogonal Recurrent Neural Networks with Scaled
               {Cayley} Transform},
  author    = {Helfrich, Kyle E. and Willmott, Devin and Ye, Qiang},
  booktitle = {Proc.\ International Conference on Machine Learning (ICML)},
  pages     = {1970--1978},
  year      = {2018},
  url       = {https://arxiv.org/abs/1707.09520}
}

@inproceedings{trockman2021cayley,
  title     = {Orthogonalizing Convolutional Layers with the
               {Cayley} Transform},
  author    = {Trockman, Asher and Kolter, J. Zico},
  booktitle = {Proc.\ International Conference on Learning
               Representations (ICLR)},
  year      = {2021},
  url       = {https://arxiv.org/abs/2104.07167}
}

@inproceedings{vaswani2017attention,
  title     = {Attention Is All You Need},
  author    = {Vaswani, Ashish and Shazeer, Noam and Parmar, Niki and
               Uszkoreit, Jakob and Jones, Llion and Gomez, Aidan N. and
               Kaiser, {\L}ukasz and Polosukhin, Illia},
  booktitle = {Advances in Neural Information Processing Systems (NeurIPS)},
  pages     = {5998--6008},
  year      = {2017},
  url       = {https://arxiv.org/abs/1706.03762}
}

@inproceedings{ba2016layernorm,
  title     = {Layer Normalization},
  author    = {Ba, Jimmy Lei and Kiros, Jamie Ryan and Hinton, Geoffrey E.},
  booktitle = {Proc.\ NeurIPS Workshop on Deep Learning Symposium},
  year      = {2016},
  url       = {https://arxiv.org/abs/1607.06450}
}

@inproceedings{lee2019settransformer,
  title     = {Set Transformer: A Framework for Attention-based
               Permutation-Invariant Neural Networks},
  author    = {Lee, Juho and Lee, Yoonho and Kim, Jungtaek and
               Kosiorek, Adam R. and Choi, Seungjin and Teh, Yee Whye},
  booktitle = {Proc.\ International Conference on Machine Learning (ICML)},
  pages     = {3744--3753},
  year      = {2019},
  url       = {https://arxiv.org/abs/1810.00825}
}

@inproceedings{loshchilov2019adamw,
  title     = {Decoupled Weight Decay Regularization},
  author    = {Loshchilov, Ilya and Hutter, Frank},
  booktitle = {Proc.\ International Conference on Learning Representations (ICLR)},
  year      = {2019},
  url       = {https://arxiv.org/abs/1711.05101}
}

@inproceedings{chopra2005contrastive,
  title     = {Learning a Similarity Metric Discriminatively, with Application
               to Face Verification},
  author    = {Chopra, Sumit and Hadsell, Raia and LeCun, Yann},
  booktitle = {Proc.\ IEEE Conference on Computer Vision and Pattern
               Recognition (CVPR)},
  pages     = {539--546},
  year      = {2005},
  organization = {IEEE},
  doi       = {10.1109/CVPR.2005.202}
}

\IEEEpubidadjcol 

\end{document}